# Detection of DNA and Poly-L-Lysine using CVD Graphene-channel FET Biosensors

Aniket Kakatkar[1], T. S. Abhilash[2], R. De Alba[2], J. M. Parpia[2], and H. G. Craighead[1,a]

[1]School of Applied and Engineering Physics, Cornell University, Ithaca, New York 14853, USA
[2]Department of Physics, Cornell University, Ithaca, New York 14853, USA

A graphene channel field-effect biosensor is demonstrated for detecting the binding of double-stranded DNA and poly-l-lysine. Sensors consist of CVD graphene transferred using a clean, etchant-free transfer method. The presence of DNA and poly-l-lysine are detected by the conductance change of the graphene transistor. A readily measured shift in the Dirac Voltage (the voltage at which the graphene's resistance peaks) is observed after the graphene channel is exposed to solutions containing DNA or poly-l-lysine. The "Dirac voltage shift" is attributed to the binding/unbinding of charged molecules on the graphene surface. The polarity of the response changes to positive direction with poly-l-lysine and negative direction with DNA. This response results in detection limits of 8 pM for 48.5 kbp DNA and 11 pM for poly-l-lysine. The biosensors are easy to fabricate, reusable and are promising as sensors of a wide variety of charged biomolecules.

———
[a]hgc1@cornell.edu

Nanoscale biosensors based on Field Effect Transistors (FET) of carbon nanotubes[1], semiconductor heterostructures[2], nanowires[3], and graphene[4,5] are being actively studied. Graphene, a single planar sheet of carbon atoms with remarkable electrical properties, has attracted great attention[6]. Numerous biomolecules adsorb onto the graphene surface non-covalently. Use of graphene as a biosensor has been demonstrated in a variety of ways, including observing changes in the conduction of the graphene[4,5,7-11]. Most devices use exfoliated graphene[4,5,12,13]. Although exfoliation yields the highest graphene quality, this process is not scalable. Moreover, the resulting monolayer films are only a few micrometers in size, limiting their active surface area.

Graphene grown through chemical vapor deposition (CVD) is an alternative for scaling to arbitrarily large device dimensions[14]. Use of CVD-grown graphene enables an increase in sensor size, thereby minimizing the baseline noise level, which scales inversely with the square root of the sensing area[15]. However, during device fabrication, the conventional transfer process involves chemical etching of the metal growth substrate, which leads to degradation of the graphene's electrical properties. There have been multiples studies on the changes in the conduction of CVD graphene due to non-covalent surface binding[7,8,16–20].

To fabricate devices, graphene is often transferred first, followed by the patterning of contacts. Patterning exposes the graphene to polymers that necessitate additional cleaning steps to achieve desirable electrical properties such as a low gate voltage for the Dirac "peak"[5,17,19]. Recent studies demonstrated high-quality CVD graphene FET (GraFET) arrays using metal as protection layer for transfer[9,21] ; the same authors have shown that a low Dirac voltage is needed for high biomolecular functionalizability of the graphene21.  In this work, we report an alternative graphene transfer approach using the "soak-and-peel" technique[22] that employs deionized (DI) water to facilitate fabrication of large sized (50 μm x 50 μm), back-gated GraFET biosensors.  In our adaptation, graphene is transferred onto a pre-patterned substrate with source-drain electrodes deposited prior to transfer. This process avoids post-transfer lithographic processing, and concomitant contamination. We observe that the processed devices exhibit Dirac voltages of about 2 V under ambient conditions. We demonstrate that GraFETs can be used for studying the surface binding of double-stranded lambda DNA (λ DNA) and poly-l-lysine.

We have fabricated GraFETs on a p-doped Si wafer which serves as the back-gate, with a gate oxide consisting of 300 nm thermally grown $SiO_2$. Using standard photo-lithographic techniques, source and drain electrodes were patterned with a channel length of 50 μm. Chromium (10 nm) and gold (40 nm) were deposited via electron-beam evaporation and lift-off was carried out. Graphene films were grown on Cu foils in a CVD[14] furnace. The graphene quality was characterized by scanning electron microscopy (SEM) and Raman microscopy.

Clean, etchant-free transfer of graphene to devices by mechanical delamination from growth substrates using a DI-water based soak-and-peel method has been demonstrated recently[22]. In this scheme, a sacrificial layer (kapton tape) is stuck to the PMMA/graphene/Cu stack and placed in hot DI-water. Water penetrates the graphene (hydrophobic) and Cu (hydrophilic) interface, separating them. The peeled-off PMMA/graphene can then be transferred to a target substrate. Though this scheme eliminates the need for wet chemical etching, re-spinning of the polymer layer and chemical solvents are still required, compromising the graphene surface quality. A significant reduction in surface residues on graphene transistor has also recently been reported using PDMS for transfer[23]. Here, we fabricate GraFETs employing the soak-and-peel delamination scheme, forgoing the PMMA layer, and using PDMS as a sacrificial layer. The PDMS/graphene/Cu stack is placed in hot DI-water (90°C) for two hours and the Cu is subsequently peeled-off. The remaining PDMS/graphene stack is pressed onto a pre-patterned substrate and heated to 140°C for 15 minutes, completing the transfer. The PDMS layer is then peeled-off (see supplementary section 1). The delamination from Cu results in a near-pristine graphene interface and improved electrical contact to substrate electrodes.

The device design and a circuit diagram used for the electrical detection of analytes is given in Fig. 1a. Fig. 1b is an optical micrograph of graphene transferred onto a pre-patterned source-drain electrode. The gating curve (source-drain current *vs.* back gate voltage) for an as-deposited device is shown in Fig. 1c. All measurements are performed with a source-drain voltage of $V_{SD}$ = 50 mV. The Dirac point is observed at ~2 V, indicating that the transfer scheme is relatively clean. The deviation from the ideal Dirac point at 0 V could result from a combination of trapped charges in the oxide and the substrate[24] as well as the graphene quality.

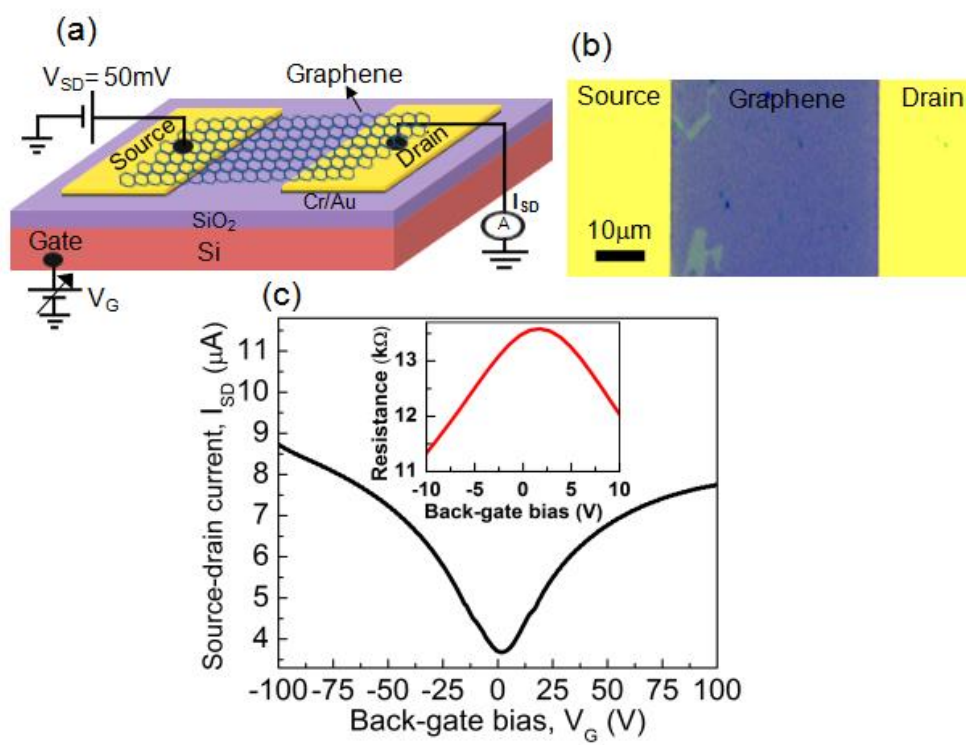

FIG. 1: a) Illustration of graphene device structure and the circuit diagram corresponding to measurement of $I_{SD}$ ($V_G$). b) Optical micrograph (false color) of graphene transferred on to a pre-patterned source-drain electrode. c) Current-gate voltage ($I_{SD}$–$V_G$) characteristic measurement of a bare GraFET. Inset shows the measured channel resistance as a function of back gate bias.

Two biomolecules have been detected with our Graphene FET (i) λ DNA obtained from New England Biolabs, and (ii) poly-l-lysine, obtained from Sigma Aldrich. Poly-l-lysine is a polymer of the essential amino acid L-lysine, with a molecular weight ~30,000 g/mol. λ DNA is double-stranded with a length of 48,502 base pairs (molecular weight of 3.15 x $10^7$ g/mol.) and is often used as a molecular weight standard in biology. Poly-l-lysine has a net positive charge, while λ DNA has a negative charge associated with it. To expose the graphene surface to different concentrations of bio-molecules, specific amounts of the poly-l-lysine and DNA are dissolved in solvent (DI-water for poly-l-lysine, Tris-EDTA (TE) buffer for λ DNA). The GraFET is dipped in solution for ~30s. That time (inset of Fig. 3b) is found to be sufficient to achieve equilibrium adsorption of analyte onto the graphene. After exposure to the mixture of solvent and analyte, the liquid on the chip is gently blown-off with nitrogen gas and the device's electrical characteristics are measured. To eliminate any shift in the Dirac voltage due to the solvent (relative to air), measurements are carried out using the solvent as reference. Post-exposure channel conductance is measured as a function of gate voltage with 50 mV source-drain voltage. The use of back gating avoids the uncertain voltage drop normally seen in solution-gated transistors and the need for insulation of contacts[16].

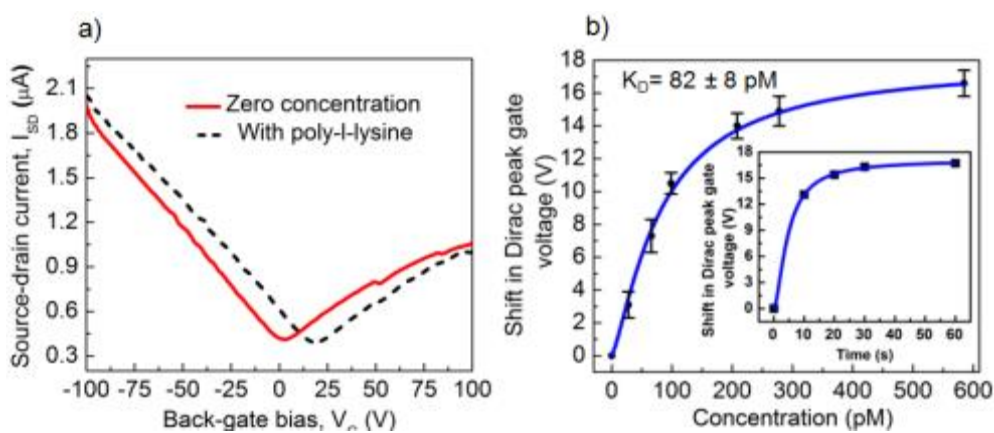

FIG. 2: a) $I_{SD}$ vs. $V_G$ for a device showing Dirac peak after exposure to water and after exposure to a ~580 pM poly-l-lysine solution in water. The Dirac voltage shifted by 17 V after exposure to the poly-l-lysine solution. b) Shift of GraFET Dirac voltage as a function of poly-l-lysine concentration relative to the Dirac voltage measured after water exposure. Error bars reflect standard error of the mean of the responses based on three different measurements on two devices. The data are fitted to the Hill equation (blue line). Inset shows the Dirac peak shift with exposure time for poly-l-lysine.

The reference gating curve (with water) for a device is measured and the conductance minimum associated with the Dirac peak is typically found at ~2 V. Upon exposure to poly-l-lysine (~580 pM), the Dirac peak shifts by 17 V as shown in Fig. 2a. To investigate the mechanism governing the binding of the biomolecules to graphene, we studied sensor response as a function of concentration. Fig. 2b shows a plot of the shift in the Dirac point (minimum conductance) with varying concentrations of poly-l-lysine. For each measurement, the GraFET is dipped in the appropriate concentration of the biomolecule for 30s, gently blow dried and the Dirac point measured. The binding follows the Langmuir adsorption model (blue curve in Fig. 2b). Following each measurement, the device is rinsed thoroughly in running water, and dried. We then confirmed that the device returns to its initial characteristic (Dirac voltage ~2 V) before taking subsequent measurements.

We extracted $K_D$, the dissociation constant for poly-l-lysine-graphene binding by fitting the concentration dependent voltage shift data to the Hill equation

$$V_{eqm} = V_{max} * \frac{1}{1 + \left(\frac{K_D}{[A]}\right)^h}$$

where $V_{eqm}$ and $V_{max}$ are equilibrium and maximum shifts in the voltage, $h$ is fitting parameter and $[A]$ is the concentration of the biomolecule to be detected. The Hill equation is equivalent to the Langmuir model when h=1. The value of $K_D$ for poly-l-lysine on graphene is found to be 82 ± 8 pM and $h$ is 1.1 ± 0.3.

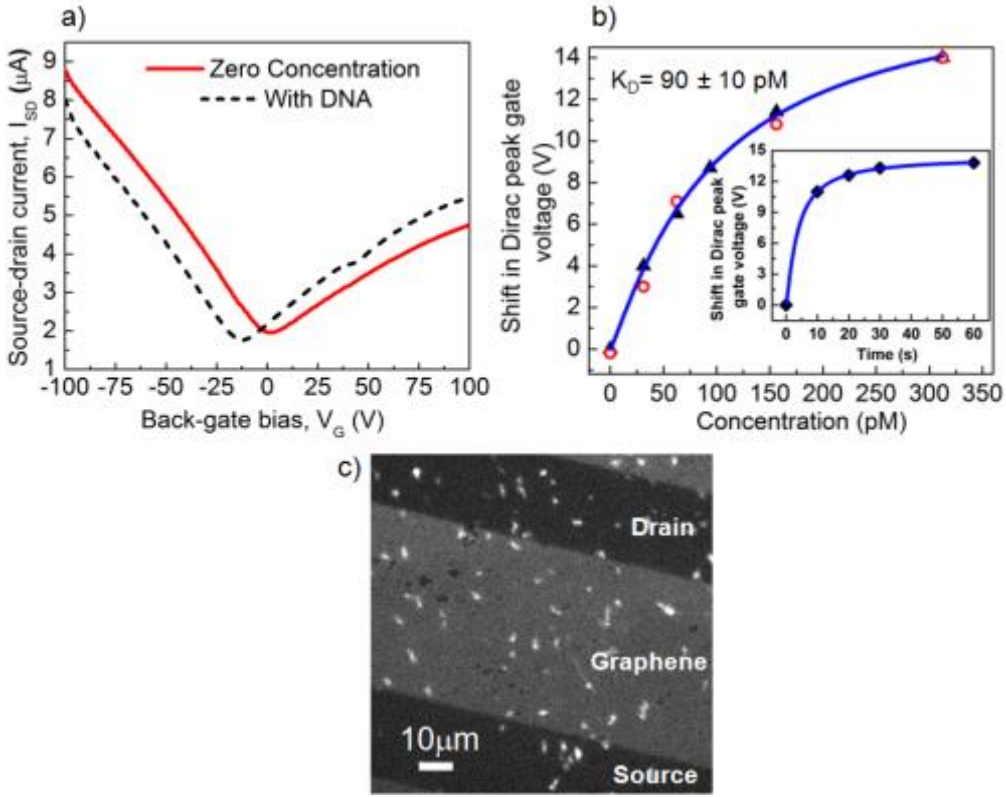

FIG. 3: a) Transistor curve for a device measured after exposure to a 300 pM solution of λ DNA. The Dirac point is close to zero after dipping in TE buffer, and shifts to negative voltages after exposure with DNA. b) The effective shift in the Dirac peak is plotted as a function of different concentrations of λ DNA. The data are fitted using the Hill equation (blue line). The measurements are done in both directions by dipping the device sequentially from low to high concentration and vice versa. Inset

shows the Dirac peak shift with exposure time for λDNA. c) Fluorescence micrograph of graphene with YOYO-1 labeled λ DNA. Bright spots indicate the presence of DNA.

Fig. 3a shows the data for detection of λ DNA. The Dirac peak shifts by -14 V upon exposure to a 300 pM λ DNA concentration. Fig. 3b shows the concentration dependent binding curve for λ DNA following an experimental procedure similar to that used for poly-l-lysine (with TE buffer as solvent instead of water). Triangles and circles indicate the measurement sequence from low to high and high to low concentrations, respectively. The measured shift in the Dirac voltage at a particular concentration is nearly identical (irrespective of whether data is acquired while the concentration is increased or decreased), lending evidence to a non-covalent mechanism governing this change. The data are fitted to the Hill equation (blue line) and yields a $K_D$ value of 90 ± 10 pM and $h$ of 1.1 ± 0.1. Electrostatic gating has been used to explain the shift of Dirac peak upon binding of charged biomolecules on graphene[5,21]. Both positive and negative shifts of the Dirac peak have been observed for single and double stranded DNA in different conditions.[5,11,20,21].

Another confirmation of the adsorption of λ DNA onto graphene was obtained through fluorescence microscopy (Fig. 3c). Images were taken on a GraFET device before and after binding. DNA is seen as ~2-3 μm sized bright spots on the graphene surface. Some of the adsorbed molecules become elongated–possibly due to the blow-drying process. Upon rinsing under running water and drying, the DNA entirely desorbs from the graphene (see supplementary section 2). The biosensors have a detection limit of 11 pM for poly-l-lysine and 8 pM for λ DNA. This is calculated from three times the standard deviation[25] of the Dirac peak voltage shift at zero concentration, which are 1.4 V and 1.2 V for poly-l-lysine and DNA respectively.

In conclusion, we have developed CVD-graphene-based bioelectronic sensors following a simple, clean transfer technique. Sensors show large shifts in the Dirac voltage when exposed to poly-l-lysine or λ DNA. The Dirac peak shifts by 17 V after exposure to ~580 pM of poly-l-lysine and by 14 V upon exposure to 300 pM of DNA. The present results indicate that binding can be described by Langmuir adsorption. Our back-gated GraFET devices are recyclable and reusable without performance degradation. This fabrication scheme can be utilized for fabricating graphene field effect sensor arrays.

Acknowledgements

We thank the Cornell NanoScale Science and Technology Facility. We would like to acknowledge Peter Rose for helping with CVD graphene growth, and Jia-Wei Yeh for fluorescence microscope imaging. We acknowledge the financial support from the Cornell Center for Materials Research under DMR 1120298 and by the NSF under DMR1202991.